\documentclass[useAMS,usenatbib]{mn2e}
\usepackage[dvips]{graphicx}

\title[Searching for the orbital period of the ultraluminous X-ray source NGC 1313 X-2]{Searching for the orbital period of the ultraluminous X-ray source NGC 1313 X-2}
\author[L. Zampieri, D. Impiombato, R. Falomo, F. Gris\'e and R. Soria]{L. Zampieri $^1$\thanks{E-mail: luca.zampieri@oapd.inaf.it},
D. Impiombato$^{1}$, R. Falomo$^{1}$, F. Gris\'e$^2$ and R. Soria$^3$\\
$^1$INAF-Osservatorio Astronomico di Padova, Vicolo dell'Osservatorio 5, 35122 Padova, Italy\\
$^2$Department of Physics and Astronomy, University of Iowa, Van Allen Hall, Iowa City, IA 52242, USA\\
$^3$Curtin Institute of Radio Astronomy, Curtin University, 1 Turner Ave, Bentley WA 6102, Australia}

\begin{document}

\date{Accepted ... Received ...; in original form ...}

\pagerange{\pageref{firstpage}--\pageref{lastpage}} \pubyear{2002}

\maketitle

\label{firstpage}

\begin{abstract}
We analyzed the longest phase-connected photometric dataset available for NGC 1313 X-2,
looking for the $\sim$6 day modulation reported by Liu et al. (2009). The folded $B$ band
light curve shows a 6 day periodicity with a significance slightly larger than $3\sigma$.
The low statistical significance of this modulation, along with the lack of detection in 
the $V$ band, make its identification uncertain.
\end{abstract}


\section{Introduction}
\label{sec1}

X-ray observations of nearby galaxies show a population of point-like,
off-nuclear sources with luminosities (if isotropic) in excess of the classical Eddington limit
for a 10$M_\odot$ compact object; they are referred to as Ultraluminous X-ray Sources (ULXs; 
see \citealt{f06,fs11} for a review).
Nowadays, hundreds of ULX candidates have been detected and many of them have been studied 
in detail (see e.g. \citealt{rw00,cp02,s+04,lb05,lm05,w+11,s+11}).
Their high X-ray luminosities and short term variability suggest that the majority of these
puzzling sources are likely to be accreting black holes (BHs) in binary systems (see e.g. 
\citealt{zr09} and references therein). Although X-ray data alone have provided evidence 
that ULXs are different from stellar-mass Galactic BHs -- either a different class of BHs 
or a different accretion state -- , X-ray spectral and timing
results remain consistent with various alternative scenarios, characterized by BHs of
different mass and origin. These go from the challenging intermediate-mass BHs of $\approx 10^2 - 10^4 M_\odot$ 
\citep{cm99}, to massive stellar BHs of $\simeq25-80 M_\odot$ formed from
the direct collapse of low-metallicity massive stars \citep{m+09,zr09},
to stellar-mass BHs ($\la 20 M_\odot$) accreting above the Eddington limit \citep{k+01,k08}.

A necessary pre-condition for determining (or at least constraining) the BH mass is to measure 
the period of a ULX binary system. This represents a crucial
preliminary step, required to perform an efficient spectroscopic follow-up in search of radial
velocity variations and hence measure the mass function of the system, and so the BH mass.
The best way to quantify the binary period is to study the optical emission of the counterpart
and to determine its periodic modulation. The same strategy was successfully
adopted for measuring the orbital period and mass of the compact object in Galactic BH
X-ray binaries (e.g. \citealt{vpm95}). 

To date a single dedicated monitoring campaign was performed on NGC 1313 X-2 with {\it HST}, 
providing the only tentative optical periodicity available.
NGC 1313 X-2 is located in the outskirts of the barred spiral galaxy NGC 1313 at a distance 
of 3.7–-4.27 Mpc \citep{t88,m02,r07}. Its observed X-ray luminosity varies between a few 
$\times 10^{39}$ erg/s and $\sim10^{40}$ erg/s in the 0.3–-10 keV band (e.g. \citealt{m07,fk06}).
The source has been extensively studied in the X-ray and optical bands (e.g. \citealt{z04,m07,g08} and 
references therein). The large amount of data available make this object a 
cornerstone for the study of ULXs. It belongs to a handful of ULXs clearly associated with stellar optical 
counterparts (e.g. \citealt{l04,m05,s05}). 
These optical sources appear to be almost ubiquitously hosted in young stellar environments (e.g. 
\citealt{p06,r06,l07}) and have properties consistent with those of young, 
massive stars. However, some ULXs appear to be associated with older stellar populations and 
at least one possible later type stellar counterpart is now known (\citealt{fk08,r08}),
although its spectral classification may be affected by
significant galactic and extra-galactic reddening \citep{g06}.

The optical counterpart of NGC 1313 X-2 was first identified on an ESO 3.6 m $R$ band image thanks
to Chandra's accurate astrometry, after the X-ray image was registered
on the position of SN 1978K \citep{z04}. ESO Very Large Telescope (VLT) images of the field showed
that the counterpart was actually composed of two distinct objects (C1 and C2),
separated by $\sim0.7$" \citep{m05}. Further refinement in the
astrometry and accurate modelling of the optical emission indicated that object C1 was the 
more likely counterpart \citep{l07,m07,pz08}. However, this was established 
beyond any doubt by the detection of the HeII$\lambda$4686 emission line in its optical
spectrum, a characteristic imprint of X-ray irradiation \citep{p06,g08}.
This star has an extinction-corrected absolute magnitude 
$M_B\sim-4.5$ mag and colors $(B-V)_0\sim-0.15$ mag and $(V-I)_0\sim-0.16$ mag
(\citealt{m07,g08}), consistent with a B spectral type.

The stellar environment of NGC 1313 X-2 has also provided interesting constraints. 
There are two groups of young stars spread out over $\sim200$ pc.
Isochrone fitting of the colour-magnitude diagram of these groups has been attempted and 
provides cluster ages of 20$\pm$5 Myrs \citep{p06,r06,l07,g08}.
As several other ULXs, NGC 1313 X-2 is also associated with a very extended 
($\sim$400 pc) optical emission nebula that gives important information on the 
energetics and lifetime of the system \citep{p02}. Assuming that it is
formed in one or more explosive events or that its mechanical energy comes from the ULX 
wind/jet activity, the characteristic age and energetics of the nebula turn out to be 
$\sim$1 Myr and $\approx10^{{52\div 53}}$ erg or $\sim 4\times 10^{39}$ erg/s, respectively
\citep{p06}.
The most up-to-date binary model calculation, including X-ray irradiation effects, 
finds consistency between all the available optical measurements if C1 is
a terminal-age main sequence or early giant donor of 
$12-15 M_\odot$ undergoing Roche-lobe overflow. The same calculations provide 
estimates of the masses also for the BH, which is in between 20 and $\sim100 
M_\odot$ (\citealt{pz10}; see also \cite{c+07} for a similar result including
star+disc irradiation but without considering binary evolution effects).

\begin{table*}
\label{tab1}
\centering
\caption{Log of the VLT+FORS1 and {\it HST}+WFPC2 photometric observations of the field
around NGC 1313 X-2, along with the $B$ and $V$ band differential photometry of object C1 
with respect to a reference field star (see text for details).}
\begin{tabular}{@{}ccccccccc@{}}
\hline
Obs. &  Date  &  MJD  &  Exposure  &  Telescope+Instr. &   $\Delta B$  &  $\sigma_B$  &  $\Delta V$  &  $\sigma_V$      \\
     &        &       &  (s)       &                   &               &              &              &                  \\
 \hline
 1  &    2007-10-21   & 54394.362587   & 242$\times$2 &  VLT+FORS1       & 3.140 &  0.069 &       &       \\
 2  &    2007-11-15   & 54419.220472   & 242$\times$2 &  VLT+FORS1       & 3.220 &  0.045 &       &       \\
 3  &    2007-11-15   & 54419.277624   & 242$\times$2 &  VLT+FORS1       & 3.261 &  0.076 &       &       \\
 4  &    2007-11-16   & 54420.255469   & 242$\times$2 &  VLT+FORS1       & 3.220 &  0.049 &       &       \\
 5  &    2007-12-06   & 54440.075805   & 242$\times$2 &  VLT+FORS1       & 3.172 &  0.042 &       &       \\
 6  &    2007-12-10   & 54444.173203   & 242$\times$2 &  VLT+FORS1       & 3.169 &  0.052 &       &       \\
 7  &    2007-12-14   & 54448.121299   & 242$\times$2 &  VLT+FORS1       & 3.236 &  0.071 &       &       \\
 8  &    2008-01-31   & 54496.073081   & 242$\times$2 &  VLT+FORS1       & 3.162 &  0.058 &       &       \\
 9  &    2008-03-02   & 54527.059338   & 242$\times$2 &  VLT+FORS1       & 3.215 &  0.066 &       &       \\
 10 &    2008-03-05   & 54530.053776   & 242$\times$2 &  VLT+FORS1       & 3.172 &  0.053 &       &       \\
 11 &    2008-03-08   & 54533.056052   & 242$\times$2 &  VLT+FORS1       & 3.244 &  0.065 &       &       \\ 
 12 &    2008-05-21   & 54607.911122   & 500$\times$2 &  {\it HST}+WFPC2 & 3.217  & 0.075  &        &       \\
    &    2008-05-21   & 54607.927094   & 400+700      &  {\it HST}+WFPC2 &        &        & 4.759  & 0.043 \\
 13 &    2008-05-22   & 54608.043761   & 500$\times$2 &  {\it HST}+WFPC2 & 3.098  & 0.072  &        &       \\
    &    2008-05-22   & 54608.059733   & 400+700      &  {\it HST}+WFPC2 &        &        & 4.824  & 0.045 \\
 14 &    2008-05-23   & 54609.042372   & 500$\times$2 &  {\it HST}+WFPC2 & 3.058  & 0.071  &        &       \\
    &    2008-05-23   & 54609.058344   & 400+700      &  {\it HST}+WFPC2 &        &        & 4.816  & 0.044 \\
 15 &    2008-05-24   & 54610.040983   & 500$\times$2 &  {\it HST}+WFPC2 & 3.192  & 0.074  &        &       \\
    &    2008-05-24   & 54610.056955   & 400+700      &  {\it HST}+WFPC2 &        &        & 4.841  & 0.045 \\
 16 &    2008-05-25   & 54611.038900   & 500$\times$2 &  {\it HST}+WFPC2 & 3.190  & 0.074  &        &       \\
    &    2008-05-25   & 54611.054872   & 400+700      &  {\it HST}+WFPC2 &        &        & 4.855  & 0.046 \\
 17 &    2008-05-26   & 54612.104178   & 500$\times$2 &  {\it HST}+WFPC2 & 3.241  & 0.075  &        &       \\
    &    2008-05-26   & 54612.120150   & 400+700      &  {\it HST}+WFPC2 &        &        & 4.917  & 0.047 \\
 18 &    2008-05-27   & 54613.179178   & 500$\times$2 &  {\it HST}+WFPC2 & 3.226  & 0.075  &        &       \\
    &    2008-05-27   & 54613.245150   & 400+700      &  {\it HST}+WFPC2 &        &        & 4.815  & 0.045 \\
 19 &    2008-05-28   & 54614.178483   & 500$\times$2 &  {\it HST}+WFPC2 & 3.185  & 0.074  &        &       \\
    &    2008-05-28   & 54614.243761   & 400+700      &  {\it HST}+WFPC2 &        &        & 4.877  & 0.046 \\
 20 &    2008-05-29   & 54615.177789   & 500$\times$2 &  {\it HST}+WFPC2 & 3.063  & 0.072  &        &       \\
    &    2008-05-29   & 54615.243066   & 400+700      &  {\it HST}+WFPC2 &        &        & 4.850  & 0.045 \\
 21 &    2008-05-30   & 54616.097233   & 500$\times$2 &  {\it HST}+WFPC2 & 3.110  & 0.071  &        &       \\
    &    2008-05-30   & 54616.113205   & 400+700      &  {\it HST}+WFPC2 &        &        & 4.775  & 0.043 \\ 
 22 &    2008-05-31   & 54617.109039   & 500$\times$2 &  {\it HST}+WFPC2 & 3.178  & 0.075  &        &       \\
    &    2008-05-31   & 54617.170150   & 400+700      &  {\it HST}+WFPC2 &        &        & 4.854  & 0.046 \\
 23 &    2008-06-01   & 54618.692372   & 500$\times$2 &  {\it HST}+WFPC2 & 3.326  & 0.077  &        &       \\
    &    2008-06-01   & 54618.708344   & 400+700      &  {\it HST}+WFPC2 &        &        & 4.762  & 0.043 \\ 
 24 &    2008-06-02   & 54619.173622   & 500$\times$2 &  {\it HST}+WFPC2 & 3.200  & 0.075  &        &       \\
    &    2008-06-02   & 54619.238205   & 400+700      &  {\it HST}+WFPC2 &        &        & 4.835  & 0.045 \\ 
 25 &    2008-06-03   & 54620.105567   & 500$\times$2 &  {\it HST}+WFPC2 & 3.189  & 0.074  &        &       \\
    &    2008-06-03   & 54620.166678   & 400+700      &  {\it HST}+WFPC2 &        &        & 4.788  & 0.043 \\ 
 26 &    2008-06-04   & 54621.171539   & 500$\times$2 &  {\it HST}+WFPC2 & 3.071  & 0.072  &        &       \\
    &    2008-06-04   & 54621.236122   & 400+700      &  {\it HST}+WFPC2 &        &        & 4.899  & 0.046 \\
 27 &    2008-06-05   & 54622.104178   & 500$\times$2 &  {\it HST}+WFPC2 & 3.101  & 0.072  &        &       \\
    &    2008-06-05   & 54622.164594   & 400+700      &  {\it HST}+WFPC2 &        &        & 4.888  & 0.046 \\ 
 28 &    2008-06-06   & 54623.102789   & 500$\times$2 &  {\it HST}+WFPC2 & 3.316  & 0.077  &        &       \\
    &    2008-06-06   & 54623.163206   & 400+700      &  {\it HST}+WFPC2 &        &        & 4.913  & 0.046 \\ 
 29 &    2008-06-07   & 54624.034733   & 500$\times$2 &  {\it HST}+WFPC2 & 3.419  & 0.080  &        &       \\
    &    2008-06-07   & 54624.091678   & 400+700      &  {\it HST}+WFPC2 &        &        & 4.872  & 0.045 \\ 
 30 &    2008-06-08   & 54625.100706   & 500$\times$2 &  {\it HST}+WFPC2 & 3.257  & 0.075  &        &       \\
    &    2008-06-08   & 54625.161122   & 400+700      &  {\it HST}+WFPC2 &        &        & 4.899  & 0.047 \\ 
 31 &    2008-06-09   & 54626.611817   & 500$\times$2 &  {\it HST}+WFPC2 & 3.285  & 0.075  &        &       \\
    &    2008-06-09   & 54626.627789   & 400+700      &  {\it HST}+WFPC2 &        &        & 4.833  & 0.045 \\
\hline
\end{tabular}
\end{table*}

\begin{figure*}
\includegraphics[width=150mm]{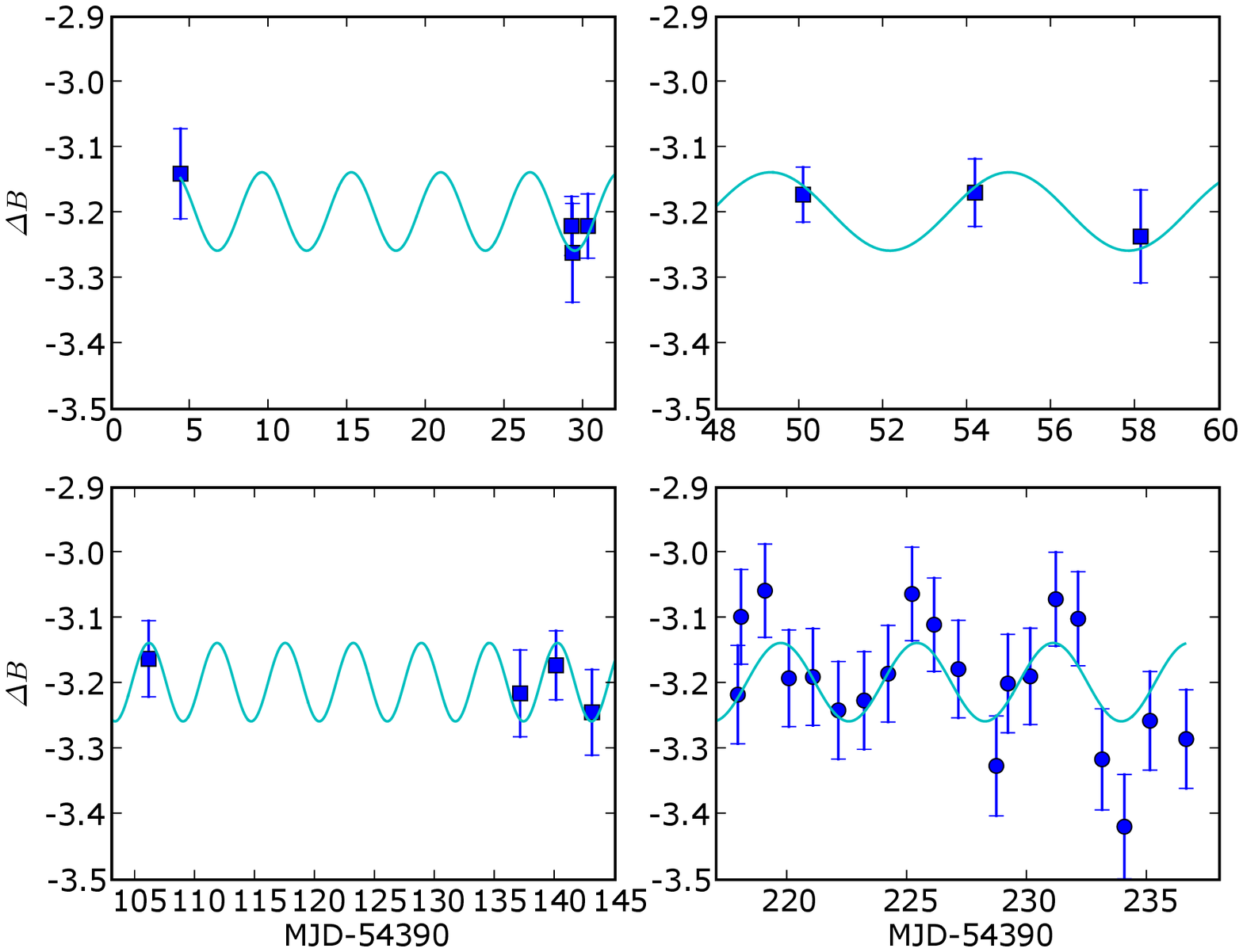}
\caption{Joint VLT+{\it HST} light curve of NGC 1313 X-2 in the $B$ band. The squares are the 
VLT+FORS1 data, while the circles represent the {\it HST}+WFPC2 observations. The magnitudes 
are the difference with those of a reference field star. The data cover a period of $\sim7.5$ months. 
The solid ({\it cyan}) line is the best fitting sinusoid with a period $P=5.68$ days.}
\label{fig1}
\end{figure*}

Quite recently, \cite{l09} found a possible periodicity of 6.12$\pm$0.16 days in a monitoring
campaign of the optical counterpart of NGC 1313 X-2 performed with the Hubble Space Telescope ({\it HST}).
This modulation was interpreted as the orbital period
of the binary system. Three cycles were detected in the $B$ band, while no modulation was found in $V$.
Previous studies carried out on the available {\it HST} and VLT observations led to negative results \citep{g08}.
More recently, lack  of significant photometric variability on a new sequence of VLT
observations has been reported by \cite{g09}.
In principle, the detection of the orbital period would definitely confirm
the identification of the optical counterpart and the binary nature of this system.
Most importantly, it would open the way to perform a dynamical measurement of 
the BH mass.  

Here we present a re-analysis of the joint VLT+FORS1 and {\it HST}+WFPC2 photometric observations 
of NGC 1313 X-2 obtained during the years 2007-2008, with the aim of clarifying the statistical
significance of the orbital periodicity identified by \cite{l09}. 
We did not consider previous VLT observations taken in 2003-2004 because they cannot be 
phase-connected to the more recent ones. In \S~\ref{sec2} we present the
data reduction procedure that we have adopted, while in \S~\ref{sec3} we show the results of the 
statistical analysis. \S~\ref{sec4} summarizes our conclusions.

\section{VLT and {\it HST} observations}
\label{sec2}

NGC 1313 X-2 was observed with VLT+FORS1 between October 2007 and March 2008 (11 epochs; 
\citealt{g09}) and with {\it HST}+WFPC2 between May and June 2008 (20 epochs; \citealt{l09}). 
During the VLT observations the sky was clear and the average seeing was in the range 0.8$''$$-$1.2$''$.
The quality of the WFPC2 images is fair, despite the degradation of the central PC chip.
A log of the observations is reported in Table~\ref{tab1}. We re-analyzed the whole dataset 
in a homogeneous way, looking for the $\sim$6 day periodicity reported by \cite{l09}.

Two exposures were taken a few minutes apart each night. For the {\it HST} dataset, we used 
the calibrated data from the WFPC2 static archive. After performing standard image 
reduction in the IRAF environment, the exposures were combined together and cleaned for cosmic rays. 
To accurately photometer the objects, we used AIDA (Astronomical Image Decomposition and 
Analysis; \citealt{uf08}), an IDL-based package originally designed to perform 
two-dimensional point-spread-function model fitting of quasar images. For the analysis of the WFPC2 exposures, 
we loaded into AIDA the appropriate point-spread-function simulated with Tiny Tim v. 
6.3\footnote{http://www.stsci.edu/software/tinytim/tinytim.html}. As the background did not vary 
significantly among the different {\it HST} exposures, we decided to keep it fixed at the average value
computed from all the observations.


\begin{table*}
\label{tab2}
\centering
\caption{Average magnitudes, standard deviation and best fitting parameters of sinusoidal fits to different 
samples of data}
\begin{tabular}{@{}cccccccc@{}}
\hline
Sample         &    $<\Delta B>$ &  $<\Delta V>$ &  $\sigma^*$   &  $P^a$    &  $A^a$  &  $\phi^a$  &  $\chi^2$(dof)  \\
               &                 &               &               &  (days) &       &          &                \\
\hline
VLT+{\it HST}$^b$ ($B$ band) & 3.198 &       & 0.078 &  $5.68_{-0.01}^{+0.01}$  &  $0.06_{-0.03}^{+0.02}$  &  ${27^0}_{-31^0}^{+29^0}$  &  21.0 (28) \\
                             &       &       &       &  $6.01_{-0.01}^{+0.01}$  &  $0.06_{-0.03}^{+0.03}$  &  ${95^0}_{-23^0}^{+25^0}$  &  21.9 (28) \\
VLT ($B$ band)           & 3.201 &       & 0.038 &   &   &   &   \\
{\it HST}     ($B$ band) & 3.196 &       & 0.094 &  $6.17_{-0.17}^{+0.18}$  &  $0.10_{-0.02}^{+0.02}$  &  ${86^0}_{-13^0}^{+13^0}$   &  13.4(17) \\
\hline
VLT+{\it HST} ($V$ band) &       & 4.843 & 0.047 &   &   &   &   \\
\hline
$^*$ Standard deviation \\
$^a$ Errors are 90\% confidence intervals. \\
$^b$ Parameters of the two deepest minima.
\end{tabular}
\end{table*}


Analyzing VLT and {\it HST} measurements together requires attention to be paid to
the systematic differences between the two photometric systems.
We then first converted the {\it HST} instrumental magnitudes 
to the standard $UBVRI$ photometric system using the updated 
transformation equations and coefficients published in Dolphin (2009) for the appropriate instrumental 
gain (which is equal to 7 in our case)\footnote{http://purcell.as.arizona.edu/wfpc2$_{-}$calib/}.
The color correction term was computed adopting the $(B-V)$ color reported in \cite{m05}.
In spite of this, residual systematic differences between the two photometric systems might still be present and
affect our measurements. In particular the color correction term is sensitive
to the overall bandpass (telescope plus atmospheric response) of the instruments. 
We then decided to perform differential photometry of the target with respect to a 
nearby field star (star D in \citealt{z04}), located on the same chip in both instruments.
The coordinates and average $BVR$ magnitudes of the optical counterpart of NGC 1313 X-2 (object C1) and 
the reference star (object D) are reported in \cite{m05}.
The reference star is brighter than the target and has a low root mean square variability ($\sim0.05$ mag
in the VLT and $\la 0.02$ mag in the {\it HST} exposures).
For similar reasons, during all the observations performed with {\it HST}+WFPC2, the field was always 
oriented in the same direction and the target and reference star were always located on the 
same position on the central PC chip. 

\begin{figure}
\includegraphics[width=80mm]{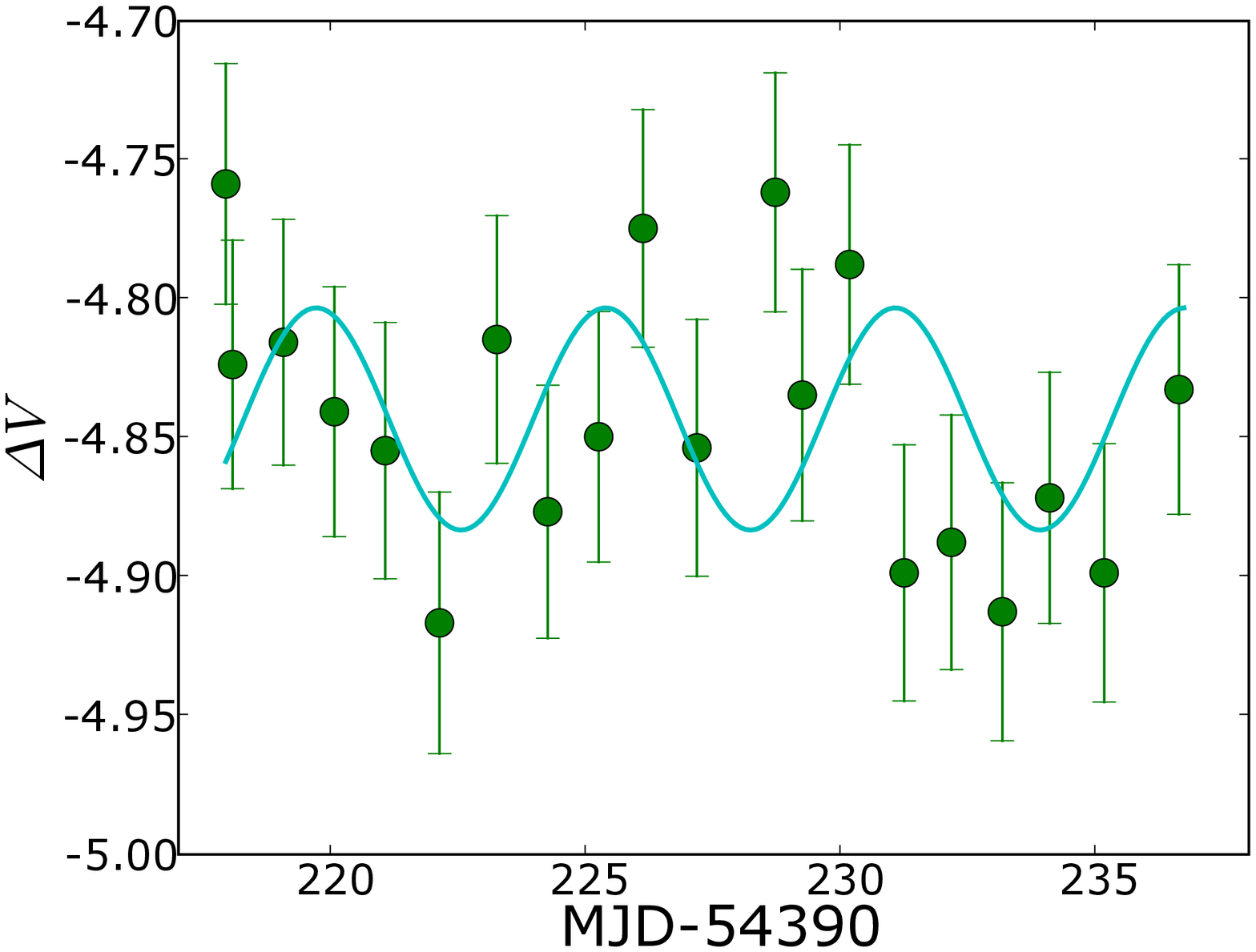}
\caption{{\it HST} light curve of NGC 1313 X-2 in the $V$ band. The magnitudes 
are the difference with those of a reference field star. The solid ({\it cyan}) line is a sinusoid with
the same period and phase obtained from the fit of the joint VLT+{\it HST} $B$ band 
data, but with a fixed amplitude of $0.04$ mag.}
\label{fig1b}
\end{figure}

Figures~\ref{fig1} and ~\ref{fig1b} show the $B$ and $V$ band light curves of NGC 1313 X-2 obtained in this way. 
The measured values of the differential magnitudes ($\Delta B=B_{\rm D}-B_{\rm C1}$, $\Delta V=V_{\rm D}-V_{\rm C1}$) 
and the corresponding errors ($\sigma_B$, $\sigma_V$) are reported in Table~\ref{tab1}.
The mean values and standard deviation of different samples of data are reported in Table~\ref{tab2}.
A comparison of our $B$ and $V$ band light curves with those of \cite{l09}
(shifted by a constant value) shows that they are consistent within the errors, 
apart from the $V$ band measurements of two observations (n. 24 and 31) that 
differ by ~0.1 mag. The same result is obtained by comparing the 
light curve of \cite{l09} with ours, before performing the $UBVRI$ 
magnitude system conversion.

Our data show clear short term ($\sim1$ day) variability.
As can be seen from Figure~\ref{fig1}, the $B$ band {\it HST} data have larger root mean square variability 
than the $B$ band VLT ones.
Similarly, comparing Figures~\ref{fig1} and ~\ref{fig1b}, it appears that
the $B$ band light curve has larger variations than the $V$ band data. 

Superimposed on these short term stochastic changes, the {\it HST} $V$ band light curve does not show significant 
regular variability (Figure~\ref{fig1b}), whereas the {\it HST} $B$ band dataset shows an approximately sinusoidal modulation with a 
period of 6 days (Figure~\ref{fig1}). The maximum peak-to-peak variation in the $B$-band light curve is 0.36 mag.


\section{Results}
\label{sec3}

Following \cite{l09}, we fitted the {\it HST} and VLT+{\it HST} $B$-band light curves
with a sinusoid:
\begin{equation}
\Delta B = <\Delta B> + A \sin (2\pi (t-t_1)/P + \phi) \, ,
\label{eq1}
\end{equation}
where $A$, $P$ and $\phi$ are the amplitude, period and phase, respectively, and $t_1=54390$
is a reference epoch. 
The best fitting parameters of the fit are reported in Table~\ref{tab2}. 
Although the values of the period and phase of the {\it HST} and VLT+{\it HST} fits are 
not consistent at the 90\% confidence level, they are in agreement at the 3-$\sigma$ level. 
Furthermore, the $\chi^2$ surface for the VLT+{\it HST} dataset
has several (at least 5) pronounced minima in the range $\sim$5.7-7.1 days
(5.7, 6.0, 6.2, 6.7, 7.1 days, respectively), with similar values of
the $\chi^2$. The values reported in Table~\ref{tab2} refer to the two
deepest minima. Considering these caveats, the fits appear to return consistent results.
The value of the amplitude and period of the {\it HST} fit are in agreement, within
the errors, with those reported by \cite{l09}.

The error on the period for the VLT+{\it HST} light curve is such that $\Delta P (T/P)\la 6$ days, where $T\sim200$ days is the interval
between the first and last observation (sampling interval). This is not larger than $P$, 
indicating that the two datasets can be phase connected. Indeed, this represents the longest 
phase-connected photometric dataset available for NGC 1313 X-2. A fit of the VLT data
alone was also attempted and returned $P=5.7$ days, $A=0.04$ mag and $\phi=28^0$.
Computing the errors for one interesting parameter (while holding the others fixed),
all these values turn out to be consistent with those reported in Table~\ref{tab2}.
However, the amplitude of the sinusoidal fit (for fixed period and phase) 
is consistent with zero at the 90\% confidence level.
Therefore, it is not possible to infer evidence of periodicity from the VLT data alone, 
but they appear to be consistent with the sinusoidal modulation observed in the {\it HST} observations. 

We checked possible systematics effects induced by the transformation from the VEGAMAG 
to the UBVRI system fitting the function in equation~(\ref{eq1}) directly to the {\it HST}
F450W band light curve. We found values for the period and amplitude completely consistent
with those of the {\it HST} $B$-band light curve reported in Table~\ref{tab2} 
($P=6.21_{-0.17}^{+0.18}$ days, $A=0.1_{-0.02}^{+0.02} $ mag). The $\chi^2$ is 31.5 for 17
d.o.f., corresponding to a null hypothesis probability of 0.0174. The rather large value
of the $\chi^2$ may indicate that, at this level of accuracy, we may be sensitive to 
small deviations of the light curve from a perfect sinusoidal shape. The phase is 
$\phi=169^0\pm 14^0$. This is clearly different from the phase reported in Table~\ref{tab2} 
because the reference epoch (MJD=54390) is $\sim 215$ days before the {\it HST} observations and
small difference in the period estimate (0.04 days) can accumulate to give rise to a phase 
shift of $\sim$0.04 days$\times (220/6)\sim 1.4$ days$\sim 80^0$.

\begin{figure}
\includegraphics[width=80mm]{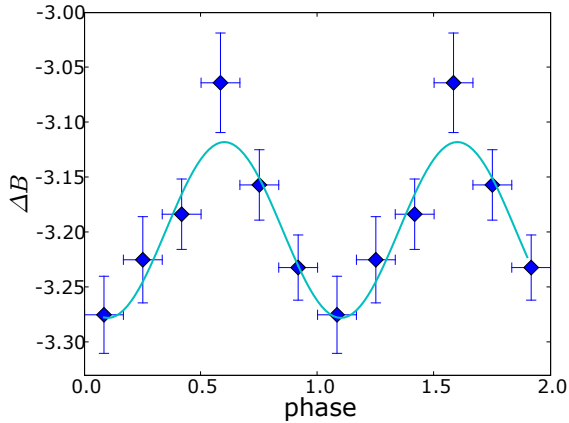}
\caption{Binned light curve (6 bins) of the $B$ band VLT+{\it HST} dataset of NGC 1313 X-2, folded 
over the best estimate of the period (6 days). The best fitting sinusoid is also shown.
The phase is measured from MJD=54390.}
\label{fig3}
\end{figure}

We tried to assess the statistical significance of the apparent $B$ band modulation in two 
different ways. First we performed a Lomb-Scargle periodogram analysis of all the unevenly
sampled time series, including both the VLT and the {\it HST} observations. The maximum power
is at frequency of $1.93\times 10^{-6}$ Hz, corresponding to a period of 6 days. However,
the null hypothesis probability is quite large (16.5\%), meaning that the statistical
significance of the modulation is low. 
In order to increase the signal-to-noise, we then decided to bin the light curve in $M=6$ bin
intervals and perform an epoch folding period search. The peak of the distribution is
$\chi^2=17.2$ at $P=6$ days. For $\nu=M-1=5$ degrees of freedom, this
corresponds to a significance level of $2.8 \sigma$. No other significant peaks were found 
in the interval between 3 and 9 days. As the validity of the $\chi^2$ distribution is limited 
to large samples, following \cite{d90} we estimated the significance of the 6 day period 
using also the $L$ statistics, which is statistically sound for all sample sizes.
The peak of the distribution is $L=6.75$ for $\nu_1=M-1=5$ and
$\nu_2=N-M=25$ degrees of freedom, where $N=31$ is the number of observations. This corresponds 
to a null hypothesis probability of $0.0004$ or a significance level of $3.6\sigma$.
In Figure~\ref{fig3} we show the folded light curve along with its best fitting sinusoid
($P=6.0$ days, $A=0.08$, $\phi=54^0$). The parameters of the sinusoid are consistent with 
those of the second best fit of the unbinned light curve (see Table~\ref{tab2}).

In contrast with the $B$ band, the $V$ band light curve shows a rather stochastic variability
(although the first 6-7 observations appear to follow a sinusoidal behaviour).
We tried a sinusoidal fit also of this dataset and found an acceptable minimum
with a periodicity $P=10.8_{-1.0}^{+1.2}$ days
that, however, is not statistically meaningful (significance from a Lomb-Scargle
periodogram analysis $\sim56\%$). As shown in Figure~\ref{fig1b}, a sinusoidal modulation 
with the same period and phase obtained from the fit of the $B$ band data and an amplitude 
$\la 0.04$ mag may be marginally consistent, within the errors, with the $V$ band data
(reduced $\chi^2 \la 1.4$).

Repeating a similar analysis on the F450W band measurements of \cite{l09}
(their Table 1), we found that the maximum power in the Lomb-Scargle periodogram
corresponds to a period $P=6.13$ days and to a null hypothesis probability of 16\%.
The peak of the $\chi^2$ distribution after binning and folding the light curve 
as described above is $\chi^2=15.8$ at $P=6.3$ days. Adopting the $L$ statistics, 
this corresponds to $L=13.8$ for $\nu_1=5$ and $\nu_2=14$ degrees of freedom,
or to a significance level of $4\sigma$ (null hypothesis probability of $0.000054$).
This value is very close to that estimated from our measurements, indicating 
that the main reason for the difference with the significance level reported by \cite{l09} 
(6$\sigma$) relies on the statistical treatment of the data and not on the different 
photometric analyses.


\section{Discussion and Conclusions}
\label{sec4}

We re-analyzed the longest phase-connected photometric dataset available for NGC 1313 X-2
using VLT+FORS1 and {\it HST}+WFPC2 observations taken between October 2007 and June 2008.
$B$-band differential photometry with respect to a nearby isolated and non-saturated 
field star confirms the 6 day modulation detected by \cite{l09} in the {\it HST}-F450W data 
alone. No significant periodic variability was found in the $V$ band. Binning the $B$ band light 
curve, the statistical significance of the 6 day modulation turns out to be slightly larger than
$3\sigma$. No other significant oscillations were found.

Although much of the work presented in this paper is a re-reduction and analysis of
previously published data \citep{l09,g09}, we emphasize that there are two important differences
in the approach that we followed here. We adopted differential photometry with respect to a 
reference field star (instead of absolute photometry) and, most importantly, we performed
a full statistical analysis of the significance of the modulation. Differential
photometry allows us to minimize the effects of absolute calibration uncertainties and
leads to small differences in the flux measurements. All but two $V$-band measurements
are in agreement with what reported previously in the literature. 
However, the major improvement of our work consists in a careful statistical re-analysis of the
data, that includes binning the light curve and using the $L$ statistics for small sample 
sizes \citep{d90}. While excluding the VLT observations from the analysis does not affect 
significantly our conclusions, we find that they can be phase-connected to the {\it HST}
dataset and are consistent with it. They also show that the optical luminosity 
of the counterpart is quite steady on timescales of months-years and that the accretion disc 
does not have a large stochastic variability, which is encouraging for photometric studies 
of this nature. At the same time, they also show that detecting periodic optical variability 
in ULX counterparts is possible both from space and from the ground, if 8+ meter class 
telescope are available. Necessary condition in the latter case is having 
optimal seeing and rather isolated counterparts.

The observed optical emission of NGC 1313 X-2 originates from the intrinsic emission 
and X-ray heating of the donor star and the accretion disc (e.g. \citealt{pz08}). 
From the maximum amplitude of the sinusoid that is marginally consistent with
the VLT data, we infer a maximum rms sinusoidal variability of 0.03 mag for the VLT data,
much smaller than that of the {\it HST} observations ($\sim0.07$ mag), indicating
that there are different levels of optical activity whose origin is unclear. This is confirmed
by the analysis of previous observations of NGC 1313 X-2 performed in 2003-2004, when the $B$ band 
light curve showed an intermediate level of variability (standard deviation $\sim$0.056 mag) with respect to the two 
datasets considered here. There may be different reasons for this behavior. 
Some B stars have intrinsic variability on a timescale of
a few hours that may be different from one cycle to the next, even though the baseline is 
constant (e.g. $\beta$ Cep stars; \citealt{s10}). This variability may be superimposed 
to a short term stochastic variability from the disc, and appear random because the timescale is too short compared to the 
time resolution of our observations. However, if we consider the standard deviation of the
residuals with respect to the sinusoidal fit, the behaviour of the VLT and {\it HST} datasets is more
similar (standard deviation $\sim$0.05 mag for VLT, $\sim$0.06 mag for {\it HST}). Therefore, it is likely
that the extra variability observed in the $B$ band {\it HST} data comes from the regular sinusoidal 
modulation. If it is so and if the changing view of the X-ray irradiated/non-irradiated donor 
surface gives a significant contribution to the modulation of the optical flux, an increased amplitude
of the modulation would imply an increment in the X-ray heating and, in turn, in the average 
optical luminosity. We know that 
the X-ray flux of NGC 1313 X-2 is indeed quite variable (e.g. \citealt{m07}). A recent Swift 
monitoring campaign of the source, performed in 2009, seems to show a bimodal distribution of 
the fluxes which may in fact suggest rapid and recurrent X-ray flares (with variations of a 
factor of 3-4; Gris\'e et al., in preparation), although the correlation with optical 
variability is not yet firmly established. 

At first glance, interpreting the extra 
{\it HST} variability as caused by increased X-ray heating does not appear to be consistent 
with the observed average $\Delta B$ reported in Table~\ref{tab2}, which is not significantly
different between the VLT and {\it HST} observations. On the other hand, as already mentioned,
it is important to remind that there may be possible residual systematic uncertainties in the 
conversion between the VEGAMAG and $UBVRI$ photometric systems. In particular, in performing
the conversion, we applied a color correction (to both objects C1 and D) estimated from the 
colors reported in \cite{m05}. However, we know that the color of object C1 is variable and
is likely to become bluer if X-ray heating increases. Furthermore, we found that the inferred $(B-V)$ 
color of the reference object D in the {\it HST} data  turns out to be slightly redder (1.6 instead of 
1.4 mag) than that reported in \cite{m05} (although they are in agreement within the errors). Both effects would
tend to change the average $\Delta B$ of the {\it HST} data and shift them upwards in 
Figure~\ref{fig1}, as expected if X-ray irradiation has increased. For an error and/or variation
in the colors of 0.15 mag, the average $\Delta B$ would vary by 0.035 mag, which is significant
in comparison with the amplitude of the modulation. Even assuming that the average luminosity
does not vary much between the two datasets, one may still think to some particular conditions under
which the modulated fraction of the emission varies without significant changes in $<\Delta B>$.
For example, when X-ray irradiation is higher, the disc may be partly obscured by material blown 
off from the inner regions. 
Or a change in the height of the outer accretion disc may partly induce
variations in the obscuration of the companion. While the total solid angle
of disc+star seen by the source (and hence the fraction of intercepted X-ray photons) 
remains constant, the relative contribution of star and disc may change, inducing variations 
in the modulated fraction of the emission. 
Clearly no definite conclusion can be reached without further accurate observations.

Also the smaller variability and absence of a detectable modulation in the $V$ band is puzzling.
Model calculations of the irradiated plus donor disc emission show that at longer 
optical wavelengths the donor spectrum declines more rapidly than the irradiated disc 
spectrum. Therefore, the contamination from the disc is comparatively stronger in the
$R$ and $V$ band with respect to the $B$ and $U$ bands \citep{pz10}. As the emission 
from the accretion disc is not orbitally modulated, any variation in the donor
star emission would induce a slightly smaller change in the $R$ and $V$ bands than in the $B$ and $U$ bands. 
Perhaps the irregular variability comes more from the disc, while the phase-modulated 
variability more from the irradiated star.
However, our analysis indicates an upper limit of 0.04 mag on the $V$ band modulation,
significantly lower than the inferred amplitude of the modulation in the $B$ band (0.09 mag).
Furthermore, ellipsoidal modulations are also likely to be important in these conditions
(e.g. \citealt{b79}), with two maxima/minima per orbital revolution. If they dominate, 
no significant wavelength dependence of the modulation is expected and the
observed 6 day periodicity would correspond to an orbital period of 12 days. 
In case both X-ray irradiation and ellipsoidal modulation contribute to
the observed variability, the combined effect may be an asymmetric light curve with
two minima of different depth (not easily detectable with the accuracy
of present measurements). However, even in this case, the light
curve may show a single maximum/minimum at superior/inferior conjunction, 
if the X-ray flux at the stellar surface is sufficiently high. Further observations 
and a detailed joint modelling of the irradiated accretion disc and ellipsoidally 
distorted donor are required to assess this point.
We note that any signature of X-ray irradiation/reprocessing in the optical
data would provide evidence against beaming, as in that case the fraction of X-ray photons 
intercepted by disc and the star would be small.

Another possibility for explaining the smaller $V$ band variability may be
the contribution of emission lines. The $B$ band contains both the HeII$\lambda$4686
and H$\beta$ lines, while the $V$ band has only the continuum. 
If irradiation and/or orbital effects are causing the lines to vary in addition to
the continuum, perhaps they may induce an extra $B$ band variability. But the
observed strength of the excess $B$ band variability ($\sim$5\% over the continuum) 
requires significant line emission and hence a very high equivalent width.
\cite{r10} found that the equivalent width of the HeII $\lambda$4686 line displays 
variations between 2 \, \AA and 11 \, \AA, not large enough to explain the observed variability.
Finally, changes in the absorbing column towards the source
may also contribute to induce differential variations 
in the $B$ and $V$ bands light curves (e.g. $E(B-V)$ variability of $\sim$0.01 mag
would cause $\Delta B$ and $\Delta V$ variability of $\sim$0.04 mag and $\sim$0.03
mag, respectively).

If the modulation is real and represents the orbital period of the ULX, it 
would place rather compelling constraints on the properties of the system and also
on the donor and BH masses. \cite{pz10} have shown that, imposing the 6 day periodicity 
and using all the other constraints available for this source from X-ray and optical observations
(mass transfer rate, position on the CM diagram, characteristic ages of the parent 
stellar cluster and the surrounding bubble nebula), the system would be consistent 
with a $\sim50-100 M_\odot$ black hole accreting from a 12--15$M_\odot$ 
star that fills its Roche lobe at terminal age main sequence. At this 
stage of the binary evolution, such a star would have a radius of $\sim 8-9\times 10^{11}$ 
cm and a separation from the BH of $\sim 4\times 10^{12}$ cm, and so it will intercept 
a fraction $\sim10-15$\% of the X-ray photons assuming isotropic emission.
Note, however, that the fraction of photons intercepted by the disc exceeds that from 
the donor star (which is also partly screened by it).

Finally, the detection of the orbital modulation would open the way to perform 
a direct dynamical measurement of the BH mass in NGC 1313 X-2, the first time ever for a ULX,
although a recent Gemini spectroscopy campaign failed to reveal regular modulations \citep{r10}.
The binned light curve suggests that the periodicity may be there, but
the low statistical significance of the $B$ band modulation, along with the lack of 
detection in the $V$ band, make its identification uncertain. A dedicated photometric 
monitoring campaign under homogeneous observing conditions to minimize systematic uncertainties 
are needed to confirm it.

\section*{Acknowledgments}
DI and LZ acknowledge financial support from INAF through grant PRIN-2007-26.
RS acknowledges financial support from a Curtin University
senior research fellowship, and hospitality at the Mullard
Space Science Laboratory (UCL) and the University of Sydney
during part of this work.
We thank the referee for his/her very careful reading of the manuscript 
and the valuable comments that helped us to improve significantly upon
the previous version of our paper. 
This paper is based on observations collected at ESO VLT.
Observations were also used from NASA/ESA {\it HST}, obtained from the data archive at the 
Space Telescope Institute. STScI is operated by the association of Universities for Research 
in Astronomy, Inc., under the NASA contract NAS 5-26555.

\label{lastpage}

\end{document}